\documentclass[a4paper,11pt]{article}
\pdfoutput=1 

\usepackage{jcappub} 

\usepackage[T1]{fontenc} 

\usepackage[T1]{fontenc} 
\usepackage{graphicx}
\usepackage{amsfonts}
\usepackage{amssymb}
\usepackage{amsbsy}
\usepackage{amsmath}
\usepackage{mathrsfs}
\usepackage{latexsym}
\usepackage{natbib}
\usepackage{bm}
\usepackage{color}
\usepackage{braket}
\usepackage{slashed}
\usepackage{pgfplots}
\usepackage{caption}

\title{Existence of conserved quantities and their algebra in curved spacetime}

\author{Susobhan Mandal}

\affiliation{Department of Physical Sciences,\\ 
Indian Institute of Science Education and Research Kolkata,\\
Mohanpur - 741 246, WB, India }

\emailAdd{sm17rs045@iiserkol.ac.in}

\abstract{In General Relativity, finding out the geodesics of a given spacetime manifold is an important task because it determines which classical processes are dynamically forbidden. Conserved quantities play an important role in solving  geodesic equations of a general spacetime manifold. Furthermore, knowing all possible conserved quantities of a system tells about the hidden symmetries of that system since, conserved quantities are deeply connected with the symmetries of the system, which are very important in their own right. Conserved quantities are also useful to capture certain features of spacetime manifold for an asymptotic observer. In this article, we show the existence of these conserved charges and their algebra for a class of dynamical systems in a generic curved spacetime.  
}

\begin{document}
\maketitle
\flushbottom
\vfill
\section{Introduction}
For a system any physical quantity $Q$ has, for each kinematically possible motion, a definite numerical value (which could be vectorial or tensorial in nature) at each instant of time. If for every dynamically allowed motion of that system it happens that $\frac{dQ}{dt}=0$, we then say that $Q$ is conserved (or that $Q$ is a constant of motion) for that particular physical system. It is often found out that the same quantity $Q$ (or a similar quantity) is a constant of motion for some broad and interesting class of systems. We then assert a conservation theorem which is in a way related to symmetries, characterizing the situations in which $Q$ is conserved.

Conservation theorems are important for various reasons. First of all, conservation theorems are generic statements about the types of motions that a dynamical law (or a class of dynamical laws) allows. In particular, they provide information about certain types of motion which are classically forbidden. Conservation theorems also give partial information about the nature of a particular motion, even if the equations are too complicated to solve analytically \cite{tito2018relativistic, grib2017comparison, liu2017geodesic, gariel2010kerr}. A conserved quantity provides a ``first integral" of the equations of motion (which is second order differential equation) and sometimes this is sufficient to essentially solve the problem; other times it can be used to decouple a set of coupled differential equations.

In General Relativity (GR), dynamics of a system is often talked \textit{w.r.t} affine parameter $\lambda$ which in certain cases can be chosen as coordinate time $t$. In an arbitrary spacetime manifold one does not expect existence of any conserved quantities however, existence of Killing symmetries in special class of spacetime manifolds, makes it possible to define certain conserved quantities easily. Furthermore, if the metric of chosen spacetime manifold is asymptotically flat then it is straightforward to assign physical meaning to such quantities (for example angular momentum, energy of a particle etc.) Another context, in which existence of conserved quantities plays an important role is to find out integrability condition of a system which basically says a 2n-dimensional hamiltonian system of ordinary differential equations (ODE) is integrable \cite{torrielli2016classical, babelon2003introduction, cao1995classical} if it has n (functionally) independent constants of the motion that are "in involution" where functionally independent means none of them can be written as a function of the others and "in involution" means that their Poisson Brackets all vanish.  

\section{Algebra of Killing vectors}

Given a spacetime manifold a general diffeomorphism is characterized by a vector field $\xi$, defined over that spacetime manifold such that under following
\begin{equation}
x'=x+\xi\implies x'^{\mu}=x^{\mu}+\xi^{\mu}(x)
\end{equation}
co-ordinate transformation, 
\begin{equation}
\delta g_{\alpha\beta}=\mathcal{L}(\xi)g_{\alpha\beta}=\nabla_{\alpha}\xi_{\beta}+\nabla_{\beta}\xi_{\alpha}=0
\end{equation}
Denoting $H_{\mu\nu}\equiv\nabla_{\mu}\xi_{\nu}+\nabla_{\nu}\xi_{\mu}$ and using Killing equation, we obtain
\begin{equation}
\begin{split}
0 & =\nabla_{\mu}H_{\nu\rho}+\nabla_{\rho}H_{\nu\mu}-\nabla_{\nu}H_{\rho\mu}\\
 & =\nabla_{\mu}\nabla_{\nu}\xi_{\rho}+\nabla_{\mu}\nabla_{\rho}\xi_{\nu}+\nabla_{\rho}\nabla_{\nu}\xi_{\mu}+\nabla_{\rho}\nabla_{\mu}\xi_{\nu}\\
 & -\nabla_{\nu}\nabla_{\rho}\xi_{\mu}-\nabla_{\nu}\nabla_{\mu}\xi_{\rho}\\
 & =[\nabla_{\mu},\nabla_{\nu}]\xi_{\rho}+[\nabla_{\rho},\nabla_{\nu}]\xi_{\mu}+\{\nabla_{\mu},\nabla_{\rho}\}\xi_{\nu}\\
 & =\mathcal{R}_{\rho\mu\nu\lambda}\xi^{\lambda}+\mathcal{R}_{\mu\rho\nu\lambda}\xi^{\lambda}+2\nabla_{\mu}\nabla_{\rho}\xi_{\nu}+\mathcal{R}_{\nu\rho\mu\lambda}\xi^{\lambda}\\
\implies & \nabla_{\mu}\nabla_{\rho}\xi_{\nu}=-\frac{1}{2}\mathcal{R}_{\nu\rho\mu\lambda}\xi^{\lambda} 
\end{split}
\end{equation}

Using this relation, it can be shown that all the isometries of a given spacetime form a Lie algebra. We show that the commutator of two isometries is another isometry
\begin{equation}
\begin{split}
\nabla_{\alpha}[\xi_{a},\xi_{b}]_{\beta} & +\nabla_{\beta}[\xi_{a},\xi_{b}]_{\alpha}=\nabla_{\alpha}(\xi_{a}^{\mu}\nabla_{\mu}\xi_{\beta}^{b}-\xi_{b}^{\mu}\nabla_{\mu}\xi_{\beta}^{a})\\
 & +\nabla_{\beta}(\xi_{a}^{\lambda}\nabla_{\lambda}\xi_{\alpha}^{b}-\xi_{b}^{\lambda}\nabla_{\lambda}\xi_{\alpha}^{a})\\
 & =\xi_{a}^{\mu}\nabla_{\alpha}\nabla_{\mu}\xi_{\beta}^{b}-\xi_{b}^{\mu}\nabla_{\alpha}\nabla_{\mu}\xi_{\beta}^{a}+\xi_{a}^{\mu}\nabla_{\beta}\nabla_{\mu}\xi_{\alpha}^{b}-\xi_{b}^{\lambda}\nabla_{\beta}\nabla_{\lambda}\xi_{\alpha}^{a}\\
 & +\nabla_{\alpha}\xi_{a}^{\mu}\nabla_{\mu}\xi_{\beta}^{b}-\nabla_{\alpha}\xi_{b}^{\mu}\nabla_{\mu}\xi_{\beta}^{a}+\nabla_{\beta}\xi_{a}^{\lambda}\nabla_{\lambda}\xi_{\alpha}^{b}-\nabla_{\beta}\xi_{b}^{\lambda}\nabla_{\lambda}\xi_{\alpha}^{a}\\
 & =-\xi_{a}^{\mu}\nabla_{\alpha}\nabla_{\beta}\xi_{\mu}^{b}+\xi_{b}^{\lambda}\nabla_{\alpha}\nabla_{\beta}\xi_{\lambda}^{a}-\xi_{a}^{\mu}\nabla_{\beta}\nabla_{\alpha}\xi_{\mu}^{b}+\xi_{b}^{\lambda}\nabla_{\beta}\nabla_{\alpha}\xi_{\lambda}^{a}\\
 & =\frac{1}{2}\xi_{a}^{\mu}\xi_{b}^{\lambda}(\mathcal{R}_{\mu\beta\alpha\lambda}-\mathcal{R}_{\lambda\beta\alpha\mu}+\mathcal{R}_{\mu\alpha\beta\lambda}-\mathcal{R}_{\lambda\alpha\beta\mu})=0
\end{split}
\end{equation}
Hence, this indicates that the mentioned commutator is a linear combination of other Killing vector with constant coefficients, known as structure constants of the isometry algebra. In general, if we consider $\{\xi_{a}, \ a=1,\ldots,r\}$ represents a basis of linear space of Killing fields of a spacetime manifold, the following mathematical statement is true
\begin{equation}\label{eqn.0.1}
[\xi_{a},\xi_{b}]=\mathcal{C}_{ab}^{c}\xi_{c}
\end{equation}
with $\mathcal{C}_{ab}^{c}=-\mathcal{C}_{ba}^{c}$. The isometry group is simply transitive if all Killing fields are linearly independent, otherwise the group is multiply transitive.

\section{Canonical formulation}
Here we discuss the Poisson bracket formalism in hamiltonian mechanics in curved spacetime. We provide here a covariant formulation of Poisson bracket which makes
it easy to understand dynamics of a given object in curved spacetime. Further, notion
of conserved charges associated to Killing symmetries is revisited and closure of their algebra is also shown in this section.
\subsection{Hamiltonian mechanics}
Most often to study dynamics of a particle on a curved spacetime one would start with following action
\begin{equation}
S=-\int\sqrt{-g_{\mu\nu}dx^{\mu}dx^{\nu}}=-\int\sqrt{-g_{\mu\nu}(x)\dot{x}^{\mu}\dot{x}^{\nu}}d\lambda
\end{equation}
where $\lambda$ is an affine parameter and $\dot{}$ represents derivative \textit{w.r.t} $\lambda$. Extremizing this action is equivalent to extremizing line-element which gives geodesic equations of that manifold.

There is another alternative choice of action which also leads to geodesic equations but characterestically of different form
\begin{equation}
S=\frac{1}{2}\int g_{\mu\nu}(x)\dot{x}^{\mu}\dot{x}^{\nu}d\lambda
\end{equation}
but both the action is equivalent from `einbin' formalism point of view. Thereore, we start with second action whose correspondence hamiltonian is
\begin{equation}\label{eqn.0}
H=\frac{1}{2}g^{\mu\nu}p_{\mu}p_{\nu}
\end{equation}
which can be obtained through Legendre transformation and where $p_{\mu}(\lambda)$ is the canonical conjugate momentum variable defined by
\begin{equation}\label{eqn.1}
p_{\mu}(\lambda)=g_{\mu\nu}(x(\lambda))\dot{x}^{\nu}
\end{equation}
which translates geodesic equations in following form
\begin{equation}\label{eqn.2}
\begin{split}
\frac{dp_{\mu}}{d\lambda} & =\partial_{\rho}g_{\mu\nu}\frac{dx^{\rho}}{d\lambda}\frac{dx^{\nu}}{d\lambda}+g_{\mu\nu}(x)\frac{d^{2}x^{\nu}}{d\lambda^{2}}\\
 & =\partial_{\rho}g_{\mu\sigma}\frac{dx^{\rho}}{d\lambda}\frac{dx^{\sigma}}{d\lambda}-\Gamma_{\mu\rho\sigma}(x)\frac{dx^{\rho}}{d\lambda}\frac{dx^{\sigma}}{d\lambda}\\
 & =\frac{dx^{\rho}}{d\lambda}\frac{dx^{\sigma}}{d\lambda}\left(\partial_{\rho}g_{\mu\sigma}-\frac{1}{2}\partial_{\rho}g_{\mu\sigma}-\frac{1}{2}\partial_{\sigma}g_{\mu\rho}+\frac{1}{2}\partial_{\mu}g_{\rho\sigma}\right)\\
 & =\Gamma_{\sigma\mu\rho}\frac{dx^{\rho}}{d\lambda}\frac{dx^{\sigma}}{d\lambda}=\Gamma_{ \ \mu}^{\sigma \ \ \rho}p_{\rho}p_{\sigma}
\end{split}
\end{equation}
The equations (\ref{eqn.1}) and (\ref{eqn.2}) constitute a pair of first order ODEs which is equivalent to second order geodesic equations. Hamilton's equations are the set of powerful results which shows how to derive above two equations from hamiltonian $H(\lambda)$ itself
\begin{equation}\label{eqn.3}
\frac{dx^{\mu}}{d\lambda}=\frac{\partial H}{\partial p_{\mu}}, \ \frac{dp_{\mu}}{d\lambda}=-\frac{\partial H}{\partial x^{\mu}}
\end{equation} 
where $H(\lambda)=\frac{1}{2}g^{\mu\nu}p_{\mu}p_{\nu}$.

\subsection{Poisson bracket}
With the definitions (\ref{eqn.3}), one can show that any function $F(x^{\mu},p_{\nu})$ defined over phase space spanned by canonical conjugate variables, changes along geodesic according to
\begin{equation}
\begin{split}
\frac{dF}{d\lambda} & =\frac{dx^{\mu}}{d\lambda}\frac{\partial F}{\partial x^{\mu}}+\frac{dp_{\mu}}{d\lambda}\frac{\partial F}{\partial p_{\mu}}\\
 & =\frac{\partial H}{\partial p_{\mu}}\frac{\partial F}{\partial x^{\mu}}-\frac{\partial H}{\partial x^{\mu}}\frac{\partial F}{\partial p_{\mu}}\\
 & =\{F,H\}
\end{split}
\end{equation} 
where Poisson bracket of two functions of phase space variable is defined as follows
\begin{equation}
\{F,G\}=-\{G,F\}=\frac{\partial F}{\partial x^{\mu}}\frac{\partial G}{\partial p_{\mu}}-\frac{\partial G}{\partial x^{\mu}}\frac{\partial F}{\partial p_{\mu}}
\end{equation}
Anti-symmetry of Poisson bracket automatically ensures that time-independent hamiltonian (which does not have explicit time dependence) is a conserved quantity since
\begin{equation}
\{H,H\}=0
\end{equation}
Poisson bracket also satisfies linearity in both of its arguments
\begin{equation}
\{F,\alpha_{1}G_{1}+\alpha_{2}G_{2}\}=\alpha_{1}\{F,G_{1}\}+\alpha_{2}\{F,G_{2}\}, \ \forall\alpha,\beta\in\mathbb{C}
\end{equation}
Another important property that it satisfies is the Jacobi identity
\begin{equation}
\{F,\{G,K\}\}+\{G,\{K,F\}\}+\{K,\{F,G\}\}=0
\end{equation}
These properties also imply that the Poisson bracket of any two conserved quantities of geodesic flow is also a conserved quantity of geodesic flow. If $F,G$ satisfies $\{F,H\}=0=\{G,H\}$ then
\begin{equation}
\{\{F,G\},H\}=\{F,\{G,H\}\}-\{G,\{F,H\}\}=0
\end{equation}

As $H$ is itself a constant of motion on geodesics the above result are enough to establish that constants of motion form a Lie-algebra, with Poisson bracket as Lie bracket. (We will alternatively use the terms constant of motion and conserved quantity which are equivalent.)

Note that definition of Poisson bracket mentioned above is nor manifestly covariant, which at first sight, seems to destroy the in buit covariance nature of GR. Since, the connetion used in GR is symmetric \cite{Wald:106274} in its lower two indices, it helps us to extend the definition of Poisson bracket, maintaining the general covariance
\begin{equation}
\{F,G\}_{*}\equiv\mathcal{D}_{\mu}F\frac{\partial G}{\partial p_{\mu}}-\frac{\partial F}{\partial p_{\mu}}\mathcal{D}_{\mu}G
\end{equation}
where for functions over phase-space variables (which are scalar by nature), we define covariant derivative as
\begin{equation}
\mathcal{D}_{\mu}F=\partial_{\mu}F+\Gamma_{ \ \mu\nu}^{\lambda}p_{\lambda}\frac{\partial F}{\partial p_{\nu}}
\end{equation}
Now note that for any two arbitrary phase-space functions
\begin{equation}
\begin{split}
\{F,G\}_{*} & =\partial_{\mu}F\frac{\partial G}{\partial p_{\mu}}-\frac{\partial F}{\partial p_{\mu}}\partial{\mu}G\\
 & +\Gamma_{ \ \mu\nu}^{\lambda}p_{\lambda}\left(\frac{\partial F}{\partial p_{\nu}}\frac{\partial G}{\partial p_{\mu}}-\frac{\partial G}{\partial p_{\nu}}\frac{\partial F}{\partial p_{\mu}}\right)\\
 & =\partial_{\mu}F\frac{\partial G}{\partial p_{\mu}}-\frac{\partial F}{\partial p_{\mu}}\partial_{\mu}G=\{F,G\}
\end{split}
\end{equation}
whose shows both the definition of Poisson bracket numerically gives same value hence, for sake of convenience we use the first definition from now onwards.

The second definition also preserves covariance manifestly for scalar functions which are of the form $J(x,p)=J^{\mu}(x)p_{\mu}$:
\begin{equation}
\begin{split}
\mathcal{D}_{\mu}J & =p_{\nu}\partial_{\mu}J^{\nu}+\Gamma_{ \ \mu\nu}^{\lambda}p_{\lambda}J^{\nu}\\
 & =(\partial_{\mu}J^{\nu}+\Gamma_{ \ \mu\lambda}^{\nu})p_{\nu}\\
 & =(\nabla_{\mu}J^{\nu})p_{\nu}
\end{split}
\end{equation}
The manifestly covariant nature of Poisson bracket can also be extended to completely symmetric tensors of rank-n by contracting all indices with $p_{\mu}$'s which yields
\begin{equation}
T(x,p)=\frac{1}{n!}T^{\mu_{1}\ldots\mu_{n}}p_{\mu_{1}}\ldots p_{\mu_{n}}
\end{equation}
A similar result can be obtained for antisymmetric tensors using Grassmann algebra.

\subsection{Conserved charges}
Symmetry of geodesic structure generated by diffeomorphisms associated with Killing vector fields implies conservation of certain quantity under geodesic flow. That quantity is obtained through contraction of Killing vector field with momentum
\begin{equation}
J[\xi]=\xi^{\mu}p_{\mu}
\end{equation}
which is also a generator of symmetry transformations.

Since this is a scalar quantity, it is invariant under coordinate transformations. It generates a coordinate transformation to which we are familiar with
\begin{equation}
\delta x^{\mu}=\{J[\xi],x^{\mu}\}=-\xi^{\mu}(x)
\end{equation}
which is nothing but the infinitesimal diffeomorphism. Similarly we can also find out the variation of momentum under this generator which is
\begin{equation}
\delta p_{\mu}=\{J[\xi],p_{\mu}\}=\partial_{\mu}\xi^{\nu}p_{\nu}
\end{equation}
which is also a coordinate dependent transformation. It can be explicitly checked that under these two transformation, hamiltonian in (\ref{eqn.0}) remains invariant which means
\begin{equation}
\frac{dJ[\xi]}{d\lambda}=\{J[\xi],H\}=0
\end{equation}
This result was expected since hamiltonian depends on the coordinates only through metric which is invariant under diffeomorphism \cite{kiriushcheva2008diffeomorphism} associated with Killing vector fields.

As we have seen that Killing equations are linear in nature and hence, it forms a linear vector space. Let choose dimension of the vector space to be $r$, then any arbitrary Killing vector field on this vector space can be written in terms of linear combination of basis vectors $\{e_{i}(x), \ i=1,\ldots,r\}$
\begin{equation}
\xi(\alpha_{i})=\alpha_{1}e_{1}(x)+\ldots\alpha_{r}e_{r}(x)
\end{equation}
Hence, each such diffeomorphism under which metric remains invariant can be written as a linear combination of $r$ numbers of conserved generators/charges whose coefficients will also be $\{\alpha_{i}\}$s
\begin{equation}
J[\xi]=\alpha_{1}J_{1}+\ldots+\alpha_{r}J_{r}
\end{equation} 
where $J_{i}(x,p)=e_{i}^{\mu}(x)p_{\mu}$. Now, we show that these generators form a Lie-algebra under Poisson bracket:
\begin{equation}
\begin{split}
\{J_{i},J_{j}\} & =\{J[e_{i}],J[e_{j}]\}\\
 & =\{e_{i}^{\mu}p_{\mu},e_{j}^{\nu}p_{\nu}\}\\
 & =e_{i}^{\mu}\{p_{\mu},e_{j}^{\nu}p_{\nu}\}+\{e_{i}^{\mu},e_{j}^{\nu}p_{\nu}\}p_{\mu}\\
 & =-e_{i}^{\mu}\partial_{\mu}e_{j}^{\nu}p_{\nu}+e_{j}^{\nu}\partial_{\nu}e_{i}^{\mu}p_{\mu}\\
 & =(e_{j}^{\nu}\partial_{\nu}e_{i}^{\mu}-e_{i}^{\nu}\partial_{\nu}e_{j}^{\mu})p_{\mu}
\end{split}
\end{equation}
In (\ref{eqn.0.1}) it is shown that the Killing vectors form a Lie-algebra, therefore,
\begin{equation}
\begin{split}
[e_{i},e_{j}] & =-(e_{j}^{\nu}\nabla_{\nu}e_{i}^{\mu}-e_{i}^{\nu}\nabla_{\nu}e_{j}^{\mu})\\
 & =-(e_{j}^{\nu}\partial_{\nu}e_{i}^{\mu}-e_{i}^{\nu}\partial_{\nu}e_{j}^{\mu})+e_{i}^{\nu}e_{j}^{\lambda}\Gamma_{ \ \nu\lambda}^{\mu}-e_{j}^{\nu}e_{i}^{\lambda}\Gamma_{ \ \nu\lambda}^{\mu}\\
 & =-(e_{j}^{\nu}\partial_{\nu}e_{i}^{\mu}-e_{i}^{\nu}\partial_{\nu}e_{j}^{\mu})\\
 & =\mathcal{C}_{ij}^{k}e_{k}
\end{split}
\end{equation} 
which shows
\begin{equation}
\{J_{i},J_{j}\}=-\mathcal{C}_{ij}^{k}e_{k}^{\mu}p_{\mu}=f_{ij}^{k}J_{k}
\end{equation}
where we denote $f_{ij}^{k}=-\mathcal{C}_{ij}^{k}$.

Since, $\{J_{i}\}$ forms a linear vector space with a closed algebra under a Lie bracket (which is Poisson bracket in this case) which is anti-symmetric and bilinear in its arguments hence, it is proven that $\{J_{i}\}$ form a Lie-algebra with structure constant $f_{ij}^{k}$.

\subsection{Conservation laws and algebra of conserved charges}
It was established in previous section that associated with each continuous symmetries generated by Killing vector fields, one can define conservation laws or conserved quantities (denoted by generators $J$) along geodesics. But the reverse statement is not true or in other words for each conserved quantity along geodesic flow there might not be a Killing symmetry, and one such example is hamiltonian of the system. This raises the next question - are there conserved quantities, higher order in momenta?

Let define a general phase-space function $J(x,p)$ on geodesic which is non-singular in momentum variables that means it is possible to express the quantity in following manner
\begin{equation}
J(x,p)=\sum_{k=0}^{\infty}\frac{1}{k!}J^{(k)\mu_{1},\ldots,\mu_{k}}p_{\mu_{1}}\ldots p_{\mu_{k}}
\end{equation}
where the expansion coefficients $J^{(k)\mu_{1},\ldots,\mu_{k}}$ are completely symmetric in the interchange of indices. Now we ask under what circumstatnces
\begin{equation}
\frac{dJ}{d\lambda}=\{J,H\}=0
\end{equation}
In the leading order, it implies
\begin{equation}
\frac{\partial J^{(0)}}{\partial x^{\mu}}=0
\end{equation}
which shows $J^{(0)}$ to a constant which can be redefined and make it zero.

At first order one would find back the Killing equation 
\begin{equation}
\nabla_{\mu}J_{\nu}^{(1)}+\nabla_{\nu}J_{\mu}^{(1)}=0
\end{equation}
because the term is linear in momentum. For the $k^{\text{th}}$ order following condition needs to be satisfied
\begin{equation}
\begin{split}
0 & =[J^{(k)\mu_{1},\ldots,\mu_{k}}p_{\mu_{1}}\ldots p_{\mu_{k}},g^{\mu\nu}p_{\mu}p_{\nu}]\\
 & =-\sum_{i=1}^{k}[J^{(k)\mu_{1},\ldots,\mu_{k}}p_{\mu_{1}}\ldots p_{\mu_{i-1}}p^{\mu_{i+1}}\ldots p_{\mu_{k}}\partial_{\mu_{i}}g^{\mu\nu}p_{\mu}p_{\nu}]\\
 & +\partial_{\mu}J^{(k)\mu_{1},\ldots,\mu_{k}}p_{\mu_{1}}\ldots p_{\mu_{k}}g^{\mu\nu}p_{\nu}+\partial_{\nu}J^{(k)\mu_{1},\ldots,\mu_{k}}p_{\mu_{1}}\ldots p_{\mu_{k}}g^{\mu\nu}p_{\mu}\\
 & =-\frac{1}{2}\sum_{i=1}^{k}[J_{\mu_{1},\ldots,\mu_{k}}^{(k) \ \mu_{i} \ \ }p^{\mu_{1}}\ldots p^{\mu_{i-1}}p^{\mu_{i+1}}\ldots p^{\mu_{k}}\partial_{\mu_{i}}g^{\mu\nu}p_{\mu}p_{\nu}]\\
 & +\partial_{\mu}J^{(k)\mu_{1},\ldots,\mu_{k}}p_{\mu_{1}}\ldots p_{\mu_{k}}g^{\mu\nu}p_{\nu}\\
 & =\sum_{i=1}^{k}[J_{\mu_{1},\ldots,\mu_{k}}^{(k) \ \mu_{i} \ \ }p^{\mu_{1}}\ldots p^{\mu_{i-1}}p^{\mu_{i+1}}\ldots p^{\mu_{k}}\Gamma_{\alpha\beta\mu_{i}}g^{\alpha\mu}g^{\beta\nu}p_{\mu}p_{\nu}]\\
 & +\partial_{\mu}J^{(k)\mu_{1},\ldots,\mu_{k}}p_{\mu_{1}}\ldots p_{\mu_{k}}g^{\mu\nu}p_{\nu}
\end{split}
\end{equation}
which can be written as follows
\begin{equation}\label{eqn.5}
\begin{split}
0 & =\sum_{i=1}^{k}[J_{\mu_{1},\ldots,\mu_{k}}^{(k) \ \beta \ \ }p^{\mu_{1}}\ldots p^{\mu_{i-1}}p_{\mu_{i+1}}\ldots p^{\mu_{k}}\Gamma_{ \ \mu\beta}^{\mu_{i}}g^{\mu\nu}p_{\mu_{i}}p_{\nu}]\\
 & +\partial_{\mu}J^{(k)\mu_{1},\ldots,\mu_{k}}p_{\mu_{1}}\ldots p_{\mu_{k}}g^{\mu\nu}p_{\nu}\\
 & =p^{\mu}p_{\mu_{1}}\ldots p_{\mu_{k}}\left(\partial_{\mu}J^{(k)\mu_{1},\ldots,\mu_{k}}+\sum_{i}\Gamma_{ \ \mu\beta}^{\mu_{i}}J^{(k)\mu_{1},\ldots,\beta,\ldots,\mu_{k}}\right)\\
 & =p^{\mu}p_{\mu_{1}}\ldots p_{\mu_{k}}\nabla_{\mu}J^{(k)\mu_{1}\ldots\mu_{k}}\\
\implies 0 & =\nabla_{(\mu}J_{\mu_{1}\ldots\mu_{k})}^{(k)} 
\end{split}
\end{equation}
These are nothing but the generalized Killing equations and solutions of these equations are called Killing tensors. Just like Killing vectors, Killing tensors also act as a generator of transformations but on phase space variables. Under action of this $x^{\mu}$ changes by
\begin{equation}
\delta x^{\mu}=\{J(x,p),x^{\mu}\}=-J^{(1)\mu}-\sum_{k=2}^{\infty}\frac{1}{(k-1)!}J^{(k)\mu\mu_{2}\ldots\mu_{k}}p_{\mu_{2}}\ldots p_{\mu_{k}}
\end{equation}
Note that these transformations are in general velocity dependent (specifically momentum dependent) for higher order generators/charges.

Closure of the algebra between charges/generators (which are constants along geodesic flow) of different order, follows from Jacobi identity of Poisson bracket
\begin{equation}\label{eqn.4}
\begin{split}
\{J^{(k)},J^{(l)}\} & =\{J^{(k)\mu_{1}\ldots\mu_{k}}p_{\mu_{1}}\ldots p_{\mu_{k}},J^{(l)\nu_{1}\ldots\nu_{l}}p_{\nu_{1}}\ldots p_{\nu_{l}}\}\\
 & =-\sum_{i=1}^{k}J^{(k)\mu_{1}\ldots\mu_{k}}p_{\mu_{1}}\ldots p_{\mu_{i-1}}\frac{\partial J^{(l)\nu_{1}\ldots\nu_{l}}}{\partial x^{\mu_{i}}}p_{\mu_{i+1}}\ldots p_{\mu_{k}}p_{\nu_{1}}\ldots p_{\nu_{l}}\\
 & +\sum_{j=1}^{l}p_{\mu_{1}}\ldots p_{\mu_{k}}J^{(l)\nu_{1}\ldots\nu_{l}}p_{\nu_{1}}\ldots p_{\nu_{j-1}}\frac{\partial J^{(k)\mu_{1}\ldots\mu_{k}}}{\partial x^{\nu_{j}}}p_{\nu_{j+1}}\ldots p_{\nu_{l}}\\
 & \equiv \tilde{J}^{(k+l-1)\sigma_{1}\ldots\sigma_{k+l-1}}p_{\mu_{1}}\ldots p_{\sigma_{k+l-1}}\\
\implies\{J^{(k)},J^{(l)}\} & \sim J^{(k+l-1)}
\end{split}
\end{equation}
Note that for $l=1$ above equation tell us
\begin{equation}
\{J^{(k)},J^{(1)}\}\sim J^{(k)}
\end{equation}
which means algebra of charge/generator of any order with charge/generator of order one (linear in momentum) is closed under Poisson bracket but it is not true for arbitrary values of $k$ and $l$.

Note also that the algebra shown in (\ref{eqn.4}) is similar to Virasoro algebra \cite{lundholm2005virasoro, moretti2003quantum} which is the algebra of infinite dimensional Killing vector space for 2-dimensional conformal field theories \cite{qualls2015lectures, ginsparg1988applied, blumenhagen2009introduction}.

\section{Hamiltonians linear in momentum}
\subsection{Introduction}
A hamiltonian quadratic in momentum naturally follows from Lagrangian which is quadratic in coordinate velocities through Legendre transformations. But in certain cases one can approximate a Hamiltonian quadratic in momentum such a way that only the liner momentum piece contained in it survives ultimately, one such example is shown here.

Consider a charged particle of charge $q$ on 2-dimensional space in a magnetic field in z-direction $\vec{B}=B_{0}\hat{z}$ and for sake of simplicity we consider non-relativistic system. In that case, hamiltonian is given by
\begin{equation}
H=\frac{1}{2m}\left(\vec{p}+\frac{q}{c}\vec{A}\right)^{2}=\frac{\vec{p}^{2}}{2m}+\frac{q}{mc}\vec{A}.\vec{p}+\frac{q^{2}}{2mc^{2}}\vec{A}.\vec{A}
\end{equation}
Now we choose circular gauge in which $\vec{A}=\frac{B_{0}}{2}(-y,x)$ which yields
\begin{equation}
H=\frac{\vec{p}^{2}}{2m}+\frac{qB_{0}}{2mc}(xp_{y}-yp_{x})+\frac{q^{2}B_{0}^{2}}{8mc^{2}}(x^{2}+y^{2})
\end{equation}
If we consider the quantity $\frac{qB_{0}}{c}$ which is inverse of some time scale, is high in magnitude then we can drop the first term (which is the kinetic term) in the hamiltonian which is valid for heavily charged astrophysical object moving around Neutron stars or in accretion disk then we obtain an effective hamiltonian which is of following form
\begin{equation}
H_{eff}=\omega_{c}(xp_{y}-yp_{x})+\frac{m\omega_{c}^{2}}{2}(x^{2}+y^{2}), \ \omega_{c}=\frac{qB_{0}}{2mc}
\end{equation}
Note that the Hamilton's equation for coordinate velocity becomes
\begin{equation}
\dot{x}=-\omega_{c}y, \ \dot{y}=\omega_{c}x 
\end{equation}
which leads to two simple harmonic oscillator equations
\begin{equation}
\ddot{x}=-\omega_{c}^{2}x, \ \ddot{y}=-\omega_{c}^{2}y
\end{equation}
which shows that particle is trapped in closed orbit \cite{kwon2018saving, rabhi2010dense, chashkina2012magnetic}. 

If $\frac{p_{i}}{m\omega_{c}x^{i}}\gg1$ then we can write the effective hamiltonian as follows
\begin{equation}
H_{eff}=\omega_{c}(xp_{y}-yp_{x})
\end{equation}

\subsection{Conserved charges and their algebra}
In this section we start with a general hamiltonian linear in momentum which is of following form
\begin{equation}
H=\zeta^{\mu}p_{\mu}
\end{equation}  
where $\zeta^{\mu}$ is a vector field defined over spacetime manifold and it's a dimensionful quantity.

As earlier let define a general phase-space function $\mathcal{K}(x,p)$ on geodesic which is non-singular in momentum variables that means it is possible to express the quantity in following manner
\begin{equation}
\mathcal{K}(x,p)=\sum_{k=0}^{\infty}\frac{1}{k!}\mathcal{K}^{(k)\mu_{1},\ldots,\mu_{k}}p_{\mu_{1}}\ldots p_{\mu_{k}}
\end{equation}
where the expansion coefficients $\mathcal{K}^{(k)\mu_{1},\ldots,\mu_{k}}$ are completely symmetric in the interchange of indices. Now we ask under what circumstatnces
\begin{equation}
\frac{d\mathcal{K}}{d\lambda}=\{\mathcal{K},H\}=0
\end{equation}

Hence, existence of a conserved charge of rank-k implies
\begin{equation}
\begin{split}
0 & =\{\mathcal{K}^{(k)},H\}=\{\mathcal{K}^{(k)\mu_{1},\ldots,\mu_{k}}p_{\mu_{1}}\ldots p_{\mu_{k}},\zeta^{\mu}p_{\mu}\}\\
 & =-\sum_{i=1}^{k}\mathcal{K}^{(k)\mu_{1},\ldots,\mu_{k}}p_{\mu_{1}}\ldots p_{\mu_{i-1}}\frac{\partial\zeta^{\nu}}{\partial x^{\mu_{i}}}p_{\nu}p_{\mu_{i+1}}\ldots p_{\mu_{k}}\\
 & +\frac{\partial\mathcal{K}^{(k)\mu_{1},\ldots,\mu_{k}}}{\partial x^{\nu}}\zeta^{\nu}p_{\mu_{1}}\ldots p_{\mu_{k}}\\
 & =-\sum_{i=1}^{k}\mathcal{K}^{(k)\mu_{1},\ldots\nu,\ldots,\mu_{k}}p_{\mu_{1}}\ldots p_{\mu_{i-1}}\frac{\partial\zeta^{\mu_{i}}}{\partial x^{\nu}}p_{\mu_{i}}p_{\mu_{i+1}}\ldots p_{\mu_{k}}\\
 & +\frac{\partial\mathcal{K}^{(k)\mu_{1},\ldots,\mu_{k}}}{\partial x^{\nu}}\zeta^{\nu}p_{\mu_{1}}\ldots p_{\mu_{k}}\\
 & =p_{\mu_{1}}\ldots p_{\mu_{k}}\left(\zeta^{\nu}\frac{\partial\mathcal{K}^{(k)\mu_{1},\ldots,\mu_{k}}}{\partial x^{\nu}}-\sum_{i=1}^{k}\mathcal{K}^{(k)\mu_{1},\ldots\nu,\ldots,\mu_{k}}\frac{\partial\zeta^{\mu_{i}}}{\partial x^{\nu}}\right)\\
 & =\mathcal{L}_{\zeta}\mathcal{K}^{(k)\mu_{1}\ldots\mu_{k}}p_{\mu_{1}}\ldots p_{\mu_{k}}\\
\implies & \mathcal{L}_{\zeta}\mathcal{K}^{(k)\mu_{1}\ldots\mu_{k}}=0  
\end{split}
\end{equation}
which says mathematically that in order to be a conserved charge/generator Lie-derivative of $\mathcal{K}^{(k)\mu_{1}\ldots\mu_{k}}$ must be zero along the vector field $\zeta^{\mu}$. These set of conditions are completely different from the earlier set of conditions in (\ref{eqn.5}). Hence, depending on the existence of solutions of above set of tensorial equation one can generate conserved charges of different rank.
One of the simplest quantities to look at are the vector fields $\{\xi^{(i)}\}$ for which the Lie-bracket $[\zeta,\xi^{(i)}]=0$ which gives set of conserved quantities $\{\xi^{(i)\mu}p_{\mu}\}$s.

Note that the algebra of charges of different rank remains same as we have shown earlier, which is 
\begin{equation}
\{\mathcal{K}^{(k)},\mathcal{K}^{(l)}\}\sim \mathcal{K}^{(k+l-1)}
\end{equation}

\section{Spinning objects}
\subsection{Introduction}
The dynamics of angular momentum and spin \cite{papapetrou2003spinning, dixon1970dynamics, wald1972gravitational, mashhoon1971particles, semerak1999spinning, hanson1974relativistic, kyrian2007spinning} of astrophysical objects plays an important role in the understanding of binary mergers and recent discovery of gravitational waves \cite{abbott2016observation, abbott2016gw151226, scientific2017gw170104, abbott2017gw170817, abbott2017gw170814} makes it possible to measure these properties in curved spacetime. Here, the discussion is started by mentioning the Poisson algebra between phase-space variables for point-like objects \cite{d2015covariant}. This is an idealization of a compact body since it neglects details of the internal structure by assigning the point-like object with overall fixed position, momentum and spin, known as the spinning-particle approximation. A large variety of models for spinning particles is found in the literature \cite{khriplovich1996gravitational, khriplovich2003equations, barausse2009hamiltonian}.
 
\subsection{Covariant phase-space structure}
In order to specify a hamiltonian dynamical system, three ingredients are required which are the phase space, identifying the dynamical degrees of freedom, the Poisson brackets between these dynamical degrees of freedoom defining a symplectic structure \cite{marmo1996symplectic, de2006classical, carosso2018geometric}. The hamiltonian generates the evolution of the system with given initial conditions by specifying a curve in the phase space passing through the initial point. The parametrization of phase-space is not unique, since changes in the parametrization can be compensated by redefining the Poisson brackets and the hamiltonian.

We start by defining spin-degrees of freedom, described by an anti-symmetric tensor $\Sigma^{\mu\nu}$
\begin{equation}
S^{\mu}=\frac{1}{2\sqrt{-g}}\varepsilon^{\mu\nu\kappa\lambda}u_{\nu}\Sigma_{\kappa\lambda}, \ Z^{\mu}=\Sigma^{\mu\nu}u_{\nu}
\end{equation}
where $u^{\mu}$ is a time-like unit vector satisfies $u^{\mu}u_{\mu}=-1$. By construction above two quantities satisfy following two conditions
\begin{equation}
Z^{\mu}u_{\mu}=0, \ \ S^{\mu}u_{\mu}=0
\end{equation}
which means they are space-like in nature.

The full set of phase-space co-ordinates of a spinning particle can be constructed using position co-ordinate $x^{\mu}$, the covariant momentum $p_{\mu}$ and the spin tensor $\Sigma_{\mu\nu}$, with antisymmetric Poisson brackets
\begin{equation}
\begin{split}
\{x^{\mu},p_{\nu}\}=\delta_{\nu}^{\mu}, & \ \ \{p_{\mu},p_{\nu}\}=\frac{1}{2}\Sigma^{\kappa\lambda}\mathcal{R}_{\kappa\lambda\mu\nu}\\
\{\Sigma^{\mu\nu},p_{\lambda}\} & =\Gamma_{ \ \lambda\kappa}^{\mu}\Sigma^{\nu\kappa}-\Gamma_{ \ \lambda\kappa}^{\nu}\Sigma^{\mu\kappa}\\
\{\Sigma^{\mu\nu},\Sigma^{\kappa\lambda}\} & = g^{\mu\kappa}\Sigma^{\nu\lambda}-g^{\mu\lambda}\Sigma^{\nu\kappa}-g^{\nu\kappa}\Sigma^{\mu\lambda}+g^{\nu\lambda}\Sigma^{\mu\kappa}
\end{split}
\end{equation}
Note that second Poisson bracket becomes trivial in a limit in which spin of the system vanishes which is consistent.

It is quite simple task to check that these brackets are indeed closed in the sense that they satisfy the Jacobi identities, hence a consistent symplectic structure is defined on the phase space. To have a well-defined dynamical system we need to complete the phase-space structure with a hamiltonian that generates proper-time evolution of the system. Here, we choose free-particle hamiltonian in (\ref{eqn.0}).

It can be shown explicitly that the chosen hamiltonian generates the following set of proper-time evolution equations \cite{d2015covariant}
\begin{equation}
\begin{split}
\dot{x}^{\mu}=\{x^{\mu},H\} & \implies p_{\mu}=g_{\mu\nu}\dot{x}^{\nu}\\
\dot{p}_{\mu}=\{p_{\mu},H\} & =\Gamma_{ \ \lambda\mu}^{\nu}p_{\nu}\dot{x}^{\lambda}+\frac{1}{2}\Sigma^{\kappa\lambda}\mathcal{R}_{\kappa\lambda\mu}^{ \ \ \ \nu}p_{\nu}\\
\dot{\Sigma}^{\mu\nu}=\{\Sigma^{\mu\nu},H\} & \implies\dot{\Sigma}^{\mu\nu}+\Gamma_{ \ \lambda\kappa}^{\mu}\dot{x}^{\lambda}\Sigma^{\kappa\nu}+\Gamma_{ \ \lambda\kappa}^{\nu}\dot{x}^{\lambda}\Sigma^{\mu\kappa}=0
\end{split}
\end{equation}
Substituting first equation into second equation one would obtain
\begin{equation}
\ddot{x}^{\mu}+\Gamma_{ \ \nu\lambda}^{\mu}\dot{x}^{\nu}\dot{x}^{\lambda}=\frac{1}{2}\Sigma^{\kappa\lambda}R_{\kappa\lambda \ \nu}^{ \ \ \mu}\dot{x}^{\nu}
\end{equation}
which reduces to familiar geodesic equation in $\Sigma=0$ limit.

\subsection{Conserved charges and their algebra}
The previous construction of conserved charges does not hold here since, the presence of $\Sigma^{\mu\nu}$ tensor makes $\{p_{\mu},p_{\nu}\}\neq 0$. Here we start by looking at conditions in order to construct conserved charges which are of following form
\begin{equation}
\mathcal{J}=\sum_{n=0}^{\infty}\mathcal{J}^{(2n)\mu_{1}\nu_{1}\mu_{2}\nu_{2}\ldots\mu_{n}\nu_{n}}(x)\Sigma_{\mu_{1}\nu_{1}}\ldots\Sigma_{\mu_{n}\nu_{n}}
\end{equation}
In order to construct conserved charges of rank-2k, following condition must be satisfied
\begin{equation}
\begin{split}
0 & =\{\mathcal{J}_{(2k)\mu_{1}\nu_{1}\mu_{2}\nu_{2}\ldots\mu_{k}\nu_{k}}\Sigma^{\mu_{1}\nu_{1}}\ldots\Sigma^{\mu_{k}\nu_{k}},g^{\mu\nu}p_{\mu}p_{\nu}\}\\
 & =\frac{\partial\mathcal{J}_{(2k)\mu_{1}\nu_{1}\ldots\mu_{k}\nu_{k}}}{\partial x^{\mu}}\Sigma^{\mu_{1}\nu_{1}}\ldots\Sigma^{\mu_{k}\nu_{k}}g^{\mu\nu}p_{\nu}\\
 & +\sum_{i=1}^{k}\mathcal{J}_{(2k)\mu_{1}\nu_{1}\ldots\mu_{k}\nu_{k}}\Sigma^{\mu_{1}\nu_{1}}\ldots\Sigma^{\mu_{i-1}\nu_{i-1}}\{\Sigma^{\mu_{i}\nu_{i}},p_{\mu}\}\Sigma^{\mu_{i+1}\nu_{i+1}}\ldots\Sigma^{\mu_{k}\nu_{k}}g^{\mu\nu}p_{\nu}\\
 & =\frac{\partial\mathcal{J}_{(2k)\mu_{1}\nu_{1}\ldots\mu_{k}\nu_{k}}}{\partial x^{\mu}}\Sigma^{\mu_{1}\nu_{1}}\ldots\Sigma^{\mu_{k}\nu_{k}}g^{\mu\nu}p_{\nu}\\
 & -\sum_{i=1}^{k}\mathcal{J}_{(2k)\mu_{1}\nu_{1}\ldots\mu_{i-1}\nu_{i-1}\lambda\nu_{i}\ldots\mu_{k}\nu_{k}}\Sigma^{\mu_{1}\nu_{1}}\ldots\Sigma^{\mu_{i-1}\nu_{i-1}}\Gamma_{ \ \mu\mu_{i}}^{\lambda}\Sigma^{\mu_{i}\nu_{i}}\Sigma^{\mu_{i+1}\nu_{i+1}}\ldots\Sigma^{\mu_{k}\nu_{k}}g^{\mu\nu}p_{\nu}\\
 & -\sum_{i=1}^{k}\mathcal{J}_{(2k)\mu_{1}\nu_{1}\ldots\mu_{i-1}\nu_{i-1}\mu_{i}\lambda\ldots\mu_{k}\nu_{k}}\Sigma^{\mu_{1}\nu_{1}}\ldots\Sigma^{\mu_{i-1}\nu_{i-1}}\Gamma_{ \ \mu\nu_{i}}^{\lambda}\Sigma^{\mu_{i}\nu_{i}}\Sigma^{\mu_{i+1}\nu_{i+1}}\ldots\Sigma^{\mu_{k}\nu_{k}}g^{\mu\nu}p_{\nu}\\
 & =\Bigg[\frac{\partial\mathcal{J}_{(2k)\mu_{1}\nu_{1}\ldots\mu_{k}\nu_{k}}}{\partial x^{\mu}}-\sum_{i=1}^{k}\left(\mathcal{J}_{(2k)\mu_{1}\nu_{1}\ldots\mu_{i-1}\nu_{i-1}\lambda\nu_{i}\ldots\mu_{k}\nu_{k}}\Gamma_{ \ \mu\mu_{i}}^{\lambda}+\mathcal{J}_{(2k)\mu_{1}\nu_{1}\ldots\mu_{i-1}\nu_{i-1}\mu_{i}\lambda\ldots\mu_{k}\nu_{k}}\Gamma_{ \ \mu\nu_{i}}^{\lambda}\right)\Bigg]\\
 & \times\Sigma^{\mu_{1}\nu_{1}}\ldots\Sigma^{\mu_{i-1}\nu_{i-1}}\Sigma^{\mu_{i}\nu_{i}}\Sigma^{\mu_{i+1}\nu_{i+1}}\ldots\Sigma^{\mu_{k}\nu_{k}}g^{\mu\nu}p_{\nu}\\
\implies & \nabla_{\mu}\mathcal{J}_{(2k)\mu_{1}\nu_{1}\ldots\mu_{k}\nu_{k}}=0
\end{split}
\end{equation}
where the transpositions of pairs $(\mu_{i}\nu_{i})$ are symmetric for $\mathcal{J}_{(2k)}$ and it is anti-symmetric in permutation between $\mu_{i}\leftrightarrow\nu_{i}, \ \forall i$.

Like previous case, here also
\begin{equation}
\{\mathcal{J}_{(2k)},\mathcal{J}_{(2l)}\}\sim\mathcal{J}_{(2k+2l-2)}
\end{equation}
which follows from the algebra
\begin{equation}
\{\Sigma^{\mu\nu},\Sigma^{\kappa\lambda}\}= g^{\mu\kappa}\Sigma^{\nu\lambda}-g^{\mu\lambda}\Sigma^{\nu\kappa}-g^{\nu\kappa}\Sigma^{\mu\lambda}+g^{\nu\lambda}\Sigma^{\mu\kappa}
\end{equation}
Like previous case, here also putting $l=1$ makes the algebra closed but not for any other arbitrary $l$ value which require inclusion of infinite number higher-rank conserved charges
\begin{equation}
\{\mathcal{J}_{(2k)},\mathcal{J}_{(2)}\}\sim\mathcal{J}_{(2k)}
\end{equation}
Next, we look at the conserved charges which are of the form $\mathcal{Q}^{(k)\mu_{1}\ldots\mu_{k}}p_{\mu_{1}}\ldots p_{\mu_{k}}$ which is rank-k and satisfy
\begin{equation}
\begin{split}
0 & =\{\mathcal{Q}^{(k)\mu_{1}\ldots\mu_{k}}p_{\mu_{1}}\ldots p_{\mu_{k}},g^{\mu\nu}p_{\mu}p_{\nu}\}\\
 & =\frac{1}{(n+1)!}\nabla_{(\mu}\mathcal{Q}_{(k)\mu_{1}\ldots\mu_{k})}p^{\mu_{1}}\ldots p^{\mu_{k}}p^{\mu}\\
 & +\sum_{i=1}^{k}\mathcal{Q}^{(k)\mu_{1}\ldots\mu_{k}}p_{\mu_{1}}\ldots p_{\mu_{i-1}}\{p_{\mu_{i}},p_{\mu}\}p_{\mu_{i+1}}\ldots p_{\mu_{k}}g^{\mu\nu}p_{\nu}\\
 & =\frac{1}{(n+1)!}\nabla_{(\mu}\mathcal{Q}_{(k)\mu_{1}\ldots\mu_{k})}p^{\mu_{1}}\ldots p^{\mu_{k}}p^{\mu}\\
 & +\frac{1}{2}\sum_{i=1}^{k}\mathcal{Q}^{(k)\mu_{1}\ldots\mu_{k}}p_{\mu_{1}}\ldots p_{\mu_{i-1}}\Sigma^{\kappa\lambda}\mathcal{R}_{\kappa\lambda\mu_{i}\mu}p_{\mu_{i+1}}\ldots p_{\mu_{k}}g^{\mu\nu}p_{\nu}\\
 & =\frac{1}{(n+1)!}\nabla_{(\mu}\mathcal{Q}_{(k)\mu_{1}\ldots\mu_{k})}p^{\mu_{1}}\ldots p^{\mu_{k}}p^{\mu}\\
 & +\frac{1}{2}\sum_{i=1}^{k}\mathcal{Q}^{(k)\mu_{1}\ldots\mu_{i-1}\mu\mu_{i+1}\ldots\mu_{k}}p_{\mu_{1}}\ldots p_{\mu_{i-1}}\Sigma^{\kappa\lambda}\mathcal{R}_{\kappa\lambda\mu\mu_{i}}p_{\mu_{i+1}}\ldots p_{\mu_{k}}g^{\mu_{i}\nu}p_{\nu}\\
 & =\frac{1}{(n+1)!}\nabla_{(\mu}\mathcal{Q}_{(k)\mu_{1}\ldots\mu_{k})}p^{\mu_{1}}\ldots p^{\mu_{k}}p^{\mu}\\
 & +\frac{1}{2}\sum_{i=1}^{k}\mathcal{Q}^{(k)\mu_{1}\ldots\mu_{i-1}\mu\mu_{i+1}\ldots\mu_{k}}p_{\mu_{1}}\ldots p_{\mu_{i-1}}p_{\mu_{i}}\Sigma^{\kappa\lambda}\mathcal{R}_{\kappa\lambda\mu}^{ \ \ \ \mu_{i}}p_{\mu_{i+1}}\ldots p_{\mu_{k}}\\
\implies & \nabla_{(\mu}\mathcal{Q}_{(k)\mu_{1}\ldots\mu_{k})}=0, \ \ \sum_{i=1}^{k}\mathcal{Q}^{(k)\mu_{1}\ldots\mu_{i-1}\mu\mu_{i+1}\ldots\mu_{k}}\mathcal{R}_{\kappa\lambda\mu}^{ \ \ \ \mu_{i}}=0 
\end{split}
\end{equation}
where the second condition in the last line implies
\begin{equation}
\sum_{i=1}^{k}\mathcal{Q}^{(k)\mu_{1}\ldots\mu_{i-1}\mu\mu_{i+1}\ldots\mu_{k}}\mathcal{R}_{\kappa\lambda\mu}^{ \ \ \ \mu_{i}}=[\nabla_{\kappa},\nabla_{\lambda}]\mathcal{Q}^{(k)\mu_{1}\ldots\mu_{k}}=0
\end{equation}
As we can see that the inclusion of non-zero spin to the system in curved spacetime adds further condition on $\mathcal{Q}_{(k)}$ in order to make it conserved quantity.

This is an important point to notice that unlike previous case here
\begin{equation}
\{\mathcal{Q}^{(k)},\mathcal{Q}^{(l)}\}\nsim\mathcal{Q}^{(k+l-1)}
\end{equation}
since in presence of non-zero spin
\begin{equation}
\{p_{\mu},p_{\nu}\}=\frac{1}{2}\Sigma^{\kappa\lambda}\mathcal{R}_{\kappa\lambda\mu\nu}
\end{equation}
which makes the Poisson-bracket between these conserved quantities not closed.

Now we are looking conserved quantities which are mixed both in $\Sigma^{\mu\nu}$s and $p_{\lambda}$s in following form
\begin{equation}
\mathcal{C}_{\mu_{1}\nu_{1}\ldots\mu_{k}\nu_{k}}^{(2k,l)\lambda_{1}\ldots\lambda_{l}}\Sigma^{\mu_{1}\nu_{1}}\ldots\Sigma^{\mu_{k}\nu_{k}}p_{\lambda_{1}}\ldots p_{\lambda_{l}}
\end{equation}
which is of rank-$(2k,l)$.

Conservation of above quantities put following conditions on these quantities
\begin{equation}
\begin{split}
0 & =\{\mathcal{C}_{\mu_{1}\nu_{1}\ldots\mu_{k}\nu_{k}}^{(2k,l)\lambda_{1}\ldots\lambda_{l}}\Sigma^{\mu_{1}\nu_{1}}\ldots\Sigma^{\mu_{k}\nu_{k}}p_{\lambda_{1}}\ldots p_{\lambda_{l}},g^{\mu\nu}p_{\mu}p_{\nu}\}\\
\implies & \nabla_{(\mu}\mathcal{C}_{\mu_{1}\nu_{1}\ldots\mu_{k}\nu_{k};|\lambda_{1}\ldots\lambda_{l})}^{(2k,l)}=0, \ \sum_{j=1}^{l}\mathcal{C}_{\mu_{1}\nu_{1}\ldots\mu_{k}\nu_{k}}^{(2k,l)\lambda_{1}\ldots\lambda_{j-1}\mu\lambda_{j+1}\ldots\lambda_{l}}\mathcal{R}_{\lambda\kappa\mu}^{ \ \ \ \lambda_{i}}=0\\
 & \nabla_{\mu}\mathcal{C}_{\mu_{1}\nu_{1}\ldots\mu_{k}\nu_{k}}^{(2k,l)\lambda_{1}\ldots\lambda_{l}}=0
\end{split}
\end{equation}
In the above set of conditions, first equation refers to covariant derivative acting on $\{\lambda_{i}\}$ indices and third equation refers to covariant derivative acts on lower indices pairwise $(\mu_{i},\nu_{i})$. 

Note that this suggests
\begin{equation}
\begin{split}
\{\mathcal{J}_{(2k)},\mathcal{J}_{(2l)}\} & \sim\mathcal{J}_{(2k+2l-2)}\\
\{\mathcal{Q}^{(k)},\mathcal{Q}^{(l)}\} & \sim\mathcal{Q}^{(k+l-1)}+\mathcal{C}^{(2,k+l-2)}\\
\{\mathcal{C}^{(2k,l)},\mathcal{C}^{(2m,n)}\} & \sim\mathcal{C}^{(2k+2m,l+n-1)}+\mathcal{C}^{(2k+2m-2,l+n)}+\mathcal{C}^{(2k+2m+2,l+n-2)}
\end{split}
\end{equation}
which form a closed algebra under Poisson bracket.

Whole procedure can be repeated again in principle for analyzing the case for hamiltonians linear in momentum.

Importance of finding such constants of motion or conserved quantities is that they are helpful in the analysis of spinning particle dynamics. An obvious such conserved quantity is the total spin
\begin{equation}
I=\frac{1}{2}g_{\mu\kappa}g_{\nu\lambda}\Sigma^{\mu\nu}\Sigma^{\kappa\lambda}=S_{\mu}S^{\mu}+Z_{\mu}Z^{\mu}
\end{equation}

\section{Spin coupled to curved spacetime}
In this section, we show the conserved quantities for a hamiltonian of a moving object whose spin is dynamically coupled to curvature of spacetime (suggested in \cite{d2015covariant}) in following way
\begin{equation}
\begin{split}
H & =H_{0}+H_{\Sigma}\\
H_{0} & =\frac{1}{2}g^{\mu\nu}p_{\mu}p_{\nu}\\
H_{\Sigma} & =\frac{\kappa}{4}\mathcal{R}_{\mu\nu\kappa\lambda}\Sigma^{\mu\nu}\Sigma^{\kappa\lambda}
\end{split}
\end{equation}
where $\kappa$ is a dimensionful quantity and strength of it is comparably small which also measures geodesic deviations of the object in the curved spacetime. In this case, conditions need to be satisfied by quantities in order to be conserved charges, become different because additional term of the hamiltonian which captures spin-curvature coupling.

Now let's start finding out the conditions for the conserved quantities of different form in this case. First, start with following kind of quantities
\begin{equation}\label{eqn.6}
\begin{split}
0 & =\{\mathcal{J}_{\mu_{1}\nu_{1}\ldots\mu_{k}\nu_{k}}^{(2k)}\Sigma^{\mu_{1}\nu_{1}}\ldots\Sigma^{\mu_{k}\nu_{k}},H_{0}+H_{\Sigma}\}\\
 & =p^{\mu}\nabla_{\mu}\mathcal{J}_{\mu_{1}\nu_{1}\ldots\mu_{k}\nu_{k}}^{(2k)}\Sigma^{\mu_{1}\nu_{1}}\ldots\Sigma^{\mu_{k}\nu_{k}}\\
 & +\{\mathcal{J}_{\mu_{1}\nu_{1}\ldots\mu_{k}\nu_{k}}^{(2k)}\Sigma^{\mu_{1}\nu_{1}}\ldots\Sigma^{\mu_{k}\nu_{k}},H_{\Sigma}\}
\end{split}
\end{equation}
Now we compute the second term in the above equation, since first term is already derived earlier.
\begin{equation}
\begin{split}
\{\mathcal{J}_{\mu_{1}\nu_{1}\ldots\mu_{k}\nu_{k}}^{(2k)} & \Sigma^{\mu_{1}\nu_{1}}\ldots\Sigma^{\mu_{k}\nu_{k}},H_{\Sigma}\}\\
 & =\frac{\kappa}{2}\{\mathcal{J}_{\mu_{1}\nu_{1}\ldots\mu_{k}\nu_{k}}^{(2k)} \Sigma^{\mu_{1}\nu_{1}}\ldots\Sigma^{\mu_{k}\nu_{k}},\Sigma^{\mu\nu}\}\mathcal{R}_{\mu\nu\kappa\lambda}\Sigma^{\kappa\lambda}\\
 & =\frac{\kappa}{2}\sum_{i=1}^{k}\mathcal{J}_{\mu_{1}\nu_{1}\ldots\mu_{k}\nu_{k}}^{(2k)} \Sigma^{\mu_{1}\nu_{1}}\ldots\Sigma^{\mu_{i-1}\nu_{i-1}}\{\Sigma^{\mu_{i}\nu_{i}},\Sigma^{\mu\nu}\}\Sigma^{\mu_{i+1}\nu_{i+1}}\ldots\Sigma^{\mu_{k}\nu_{k}}\mathcal{R}_{\mu\nu\kappa\lambda}\Sigma^{\kappa\lambda}\\
 & =\frac{\kappa}{2}\sum_{i=1}^{k}\mathcal{J}_{\mu_{1}\nu_{1}\ldots\mu_{k}\nu_{k}}^{(2k)} \Sigma^{\mu_{1}\nu_{1}}\ldots\Sigma^{\mu_{i-1}\nu_{i-1}}(g^{\mu_{i}\mu}\Sigma^{\nu_{i}\nu}-g^{\mu_{i}\nu}\Sigma^{\nu_{i}\mu}-g^{\nu_{i}\mu}\Sigma^{\mu_{i}\nu}+g^{\nu_{i}\nu}\Sigma^{\mu_{i}\mu})\\
 & \times\Sigma^{\mu_{i+1}\nu_{i+1}}\ldots\Sigma^{\mu_{k}\nu_{k}}\mathcal{R}_{\mu\nu\kappa\lambda}\Sigma^{\kappa\lambda}\\
 & =\frac{\kappa}{2}\sum_{i=1}^{k}\Bigg[\mathcal{J}_{\mu_{1}\nu_{1}\ldots\mu_{k}\nu_{k}}\mathcal{R}_{ \ \nu\kappa\lambda}^{\mu_{i}}\Sigma^{\nu_{i}\nu}+\mathcal{J}_{\mu_{1}\nu_{1}\ldots\mu_{k}\nu_{k}}\mathcal{R}_{ \ \mu\kappa\lambda}^{\mu_{i}}\Sigma^{\nu_{i}\mu}-\mathcal{J}_{\mu_{1}\nu_{1}\ldots\mu_{k}\nu_{k}}\mathcal{R}_{ \ \nu\kappa\lambda}^{\nu_{i}}\Sigma^{\mu_{i}\nu}\\
 & -\mathcal{J}_{\mu_{1}\nu_{1}\ldots\mu_{k}\nu_{k}}\mathcal{R}_{ \ \mu\kappa\lambda}^{\nu_{i}}\Sigma^{\mu_{i}\mu}\Bigg]\Sigma^{\mu_{1}\nu_{1}}\ldots\Sigma^{\mu_{i-1}\nu_{i-1}}\Sigma^{\mu_{i+1}\nu_{i+1}}\ldots\Sigma^{\mu_{k}\nu_{k}}\Sigma^{\kappa\lambda}\\
 & =\kappa\sum_{i=1}^{k}\Bigg[-\mathcal{J}_{\mu_{1}\nu_{1}\ldots\rho\nu_{i}\ldots\mu_{k}\nu_{k}}\mathcal{R}_{ \ \mu_{i}\kappa\lambda}^{\rho}-\mathcal{J}_{\mu_{1}\nu_{1}\ldots\mu_{i}\rho\ldots\mu_{k}\nu_{k}}\mathcal{R}_{ \ \nu_{i}\kappa\lambda}^{\rho}\Bigg]\Sigma^{\mu_{1}\nu_{1}}\ldots\Sigma^{\mu_{i}\nu_{i}}\ldots\Sigma^{\mu_{k}\nu_{k}}\\
 & =-\kappa\Big[[\nabla_{\kappa},\nabla_{\lambda}]\mathcal{J}_{\mu_{1}\nu_{1}\ldots\mu_{k}\nu_{k}}\Big]\Sigma^{\mu_{1}\nu_{1}}\ldots\Sigma^{\mu_{k}\nu_{k}}\Sigma^{\kappa\lambda}
\end{split}
\end{equation}
Hence, in a nutshell the condition put in (\ref{eqn.6}) becomes
\begin{equation}
\begin{split}
p^{\mu}\nabla_{\mu} & \mathcal{J}_{\mu_{1}\nu_{1}\ldots\mu_{k}\nu_{k}}^{(2k)}-\kappa\Sigma^{\kappa\lambda}\Big[[\nabla_{\kappa},\nabla_{\lambda}]\mathcal{J}_{\mu_{1}\nu_{1}\ldots\mu_{k}\nu_{k}}^{(2k)}\Big]=0, \ \forall \ p_{\mu}, \ \Sigma^{\kappa\lambda}\\
\implies & \nabla_{\mu}\mathcal{J}_{\mu_{1}\nu_{1}\ldots\mu_{k}\nu_{k}}^{(2k)}=0=[\nabla_{\kappa},\nabla_{\lambda}]\mathcal{J}_{\mu_{1}\nu_{1}\ldots\mu_{k}\nu_{k}}^{(2k)}
\end{split}
\end{equation}
Therefore, above two conditions need to be satisfied at same time in order $\mathcal{J}_{\mu_{1}\nu_{1}\ldots\mu_{k}\nu_{k}}^{(2k)}\Sigma^{\mu_{1}\nu_{1}}\ldots\Sigma^{\mu_{k}\nu_{k}}$ to be conserved charge.

There exist another class of quantities for which, the conditions need to be imposed in order to make them conserved charges, shown below.
\begin{equation}
\begin{split}
0 & =\{\mathcal{Q}^{(k)\mu_{1}\ldots\mu_{k}}p_{\mu_{1}}\ldots p_{\mu_{k}},H_{0}+H_{\Sigma}\}\\
 & =\frac{1}{(n+1)!}p^{\mu}\nabla_{(\mu}\mathcal{Q}_{\mu_{1}\ldots\mu_{k})}^{(k)}p^{\mu_{1}}\ldots p^{\mu_{k}}+\frac{1}{2}\Sigma^{\kappa\lambda}[\nabla_{\kappa},\nabla_{\lambda}]\mathcal{Q}_{\mu_{1}\ldots\mu_{k}}^{(k)}p^{\mu_{1}}\ldots p^{\mu_{k}}\\
 & +\{\mathcal{Q}^{(k)\mu_{1}\ldots\mu_{k}}p_{\mu_{1}}\ldots p_{\mu_{k}},H_{\Sigma}\}
\end{split}
\end{equation}
Here, we only need to evaluate the following quantity
\begin{equation}
\begin{split}
\{\mathcal{Q}^{(k)\mu_{1}\ldots\mu_{k}} & p_{\mu_{1}}\ldots p_{\mu_{k}},H_{\Sigma}\}=\frac{\kappa}{2}\{\mathcal{Q}^{(k)\mu_{1}\ldots\mu_{k}}p_{\mu_{1}}\ldots p_{\mu_{k}},\Sigma^{\mu\nu}\}\mathcal{R}_{\mu\nu\kappa\lambda}\Sigma^{\kappa\lambda}\\
 & =\frac{\kappa}{2}\sum_{i=1}^{k}\mathcal{Q}^{(k)\mu_{1}\ldots\mu_{k}}p_{\mu_{1}}\ldots p_{\mu_{i-1}}\{p_{\mu_{i}},\Sigma^{\mu\nu}\}p_{\mu_{i+1}}\ldots p_{\mu_{k}}\mathcal{R}_{\mu\nu\kappa\lambda}\Sigma^{\kappa\lambda}\\
 & -\frac{\kappa}{2}\sum_{i=1}^{k}\mathcal{Q}^{(k)\mu_{1}\ldots\mu_{k}}p_{\mu_{1}}\ldots p_{\mu_{i-1}}\frac{\partial\mathcal{R}_{\mu\nu\lambda\kappa}}{\partial x^{\mu_{i}}}p_{\mu_{i+1}}\ldots p_{\mu_{k}}\Sigma^{\mu\nu}\Sigma^{\kappa\lambda}\\
 & =-\frac{\kappa}{2}\sum_{i=1}^{k}\mathcal{Q}^{(k)\mu_{1}\ldots\mu_{k}}p_{\mu_{1}}\ldots p_{\mu_{i-1}}(\Gamma_{ \ \mu_{i}\rho}^{\mu}\Sigma^{\nu\rho}-\Gamma_{ \ \mu_{i}\rho}^{\nu}\Sigma^{\mu\rho})p_{\mu_{i+1}}\ldots p_{\mu_{k}}\mathcal{R}_{\mu\nu\kappa\lambda}\Sigma^{\kappa\lambda}\\
 & -\frac{\kappa}{2}\sum_{i=1}^{k}\mathcal{Q}^{(k)\mu_{1}\ldots\mu_{k}}p_{\mu_{1}}\ldots p_{\mu_{i-1}}\frac{\partial\mathcal{R}_{\mu\nu\lambda\kappa}}{\partial x^{\mu_{i}}}p_{\mu_{i+1}}\ldots p_{\mu_{k}}\Sigma^{\kappa\lambda}\Sigma^{\mu\nu}\\
=-\frac{\kappa}{2}\sum_{i=1}^{k} & \mathcal{Q}^{(k)\mu_{1}\ldots\mu_{k}}p_{\mu_{1}}\ldots p_{\mu_{i-1}}\left(\Sigma^{\mu\nu}\frac{\partial\mathcal{R}_{\mu\nu\lambda\kappa}}{\partial x^{\mu_{i}}}+\Gamma_{ \ \mu_{i}\rho}^{\mu}\Sigma^{\nu\rho}\mathcal{R}_{\mu\nu\kappa\lambda}-\Gamma_{ \ \mu_{i}\rho}^{\nu}\Sigma^{\mu\rho}\mathcal{R}_{\mu\nu\kappa\lambda}\right)p_{\mu_{i+1}}\ldots p_{\mu_{k}}\Sigma^{\kappa\lambda}\\
=-\frac{\kappa}{2}\sum_{i=1}^{k}\Bigg[ & \mathcal{Q}^{(k)\mu_{1}\ldots\mu_{k}}p_{\mu_{1}}\ldots p_{\mu_{i-1}}\left(\frac{\partial(\Sigma^{\mu\nu}\mathcal{R}_{\mu\nu\lambda\kappa})}{\partial x^{\mu_{i}}}-\Gamma_{ \ \mu_{i}\rho}^{\mu}\Sigma^{\rho\nu}\mathcal{R}_{\mu\nu\kappa\lambda}-\Gamma_{ \ \mu_{i}\rho}^{\nu}\Sigma^{\mu\rho}\mathcal{R}_{\mu\nu\kappa\lambda}\right)\\
 & \times p_{\mu_{i+1}}\ldots p_{\mu_{k}}\Sigma^{\kappa\lambda}\Bigg]
\end{split}
\end{equation}
Let's define an anti-symmetric rank-2 tensor $\tilde{\Sigma}_{\kappa\lambda}\equiv\mathcal{R}_{\mu\nu\kappa\lambda}\Sigma^{\mu\nu}$, then
\begin{equation}
\begin{split}
\implies\{\mathcal{Q}^{(k)\mu_{1}\ldots\mu_{k}} & p_{\mu_{1}}\ldots p_{\mu_{k}},H_{\Sigma}\}\\
=-\frac{\kappa}{2}\sum_{i=1}^{k}\Bigg[ & \mathcal{Q}^{(k)\mu_{1}\ldots\mu_{k}}p_{\mu_{1}}\ldots p_{\mu_{i-1}}\left(\Sigma^{\kappa\lambda}\frac{\partial\tilde{\Sigma}_{\kappa\lambda}}{\partial x^{\mu_{i}}}-\Gamma_{ \ \mu_{i}\rho}^{\mu}\Sigma^{\rho\nu}\tilde{\Sigma}_{\mu\nu}-\Gamma_{ \ \mu_{i}\rho}^{\nu}\Sigma^{\mu\rho}\tilde{\Sigma}_{\mu\nu}\right)p_{\mu_{i+1}}\ldots p_{\mu_{k}}\Bigg]\\
=-\frac{\kappa}{2}\sum_{i=1}^{k}\Bigg[ & \mathcal{Q}^{(k)\mu_{1}\ldots\mu_{k}}p_{\mu_{1}}\ldots p_{\mu_{i-1}}(\Sigma^{\mu\nu}\nabla_{\mu_{i}}\tilde{\Sigma}_{\mu\nu})p_{\mu_{i+1}}\ldots p_{\mu_{k}}\Bigg]\\
=-\frac{\kappa}{2}p_{\mu_{1}}\ldots p_{\mu_{k-1}} & \Bigg[\mathcal{Q}^{(k)\lambda\mu_{1}\ldots\mu_{k-1}}\Sigma^{\mu\nu}\nabla_{\lambda}\tilde{\Sigma}_{\mu\nu}+\ldots+\mathcal{Q}^{(k)\mu_{1}\ldots\mu_{k-1}\lambda}\Sigma^{\mu\nu}\nabla_{\lambda}\tilde{\Sigma}_{\mu\nu}\Bigg]\\
=-\frac{\kappa}{2}p_{\mu_{1}}\ldots p_{\mu_{k-1}} & \left(\mathcal{Q}^{(k)\lambda\mu_{1}\ldots\mu_{k-1}}+\ldots+\mathcal{Q}^{(k)\mu_{1}\ldots\mu_{k-1}\lambda}\right)\Sigma^{\mu\nu}\nabla_{\lambda}\tilde{\Sigma}_{\mu\nu}
\end{split}
\end{equation}
Therefore, altogether we have 3 conditions, need to be imposed
\begin{equation}
\begin{split}
\nabla_{(\mu}\mathcal{Q}_{\mu_{1}\ldots\mu_{k})}^{(k)}=0, & \ [\nabla_{\kappa},\nabla_{\lambda}]\mathcal{Q}_{\mu_{1}\ldots\mu_{k}}^{(k)}=0\\
\mathcal{Q}^{(k)(\lambda\mu_{1}\ldots\mu_{k-1})} & \nabla_{\lambda}\tilde{\Sigma}_{\mu\nu}\Sigma^{\mu\nu}=0
\end{split}
\end{equation}
in order to $\mathcal{Q}^{(k)\mu_{1}\ldots\mu_{k}}p_{\mu_{1}}\ldots p_{\mu_{k}}$ be  conserved charge. There is another kind of conserved charges which are mixed in momentum and $\Sigma^{\mu\nu}$. Those can also be obtained in similar way using the above mathematical procedure.

\section{Discussion}
In Newtonian mechanics, an isolated system of two point particles interacting under gravity is exactly solvable and the resulting motion is periodic. The energy and angular momentum are represented by two conserved integrals of motion which is a celebrated result in physics community. The dynamics of binaries are nonlinear in nature in General Relativity due to its inherent construction and hence, it is important to find the conserved quantities in order to predict the dynamics or to extract information about dynamics in curved spacetime. Conserved quantities also play a huge role in numerical relativity in terms of generating new solutions for example, by adding a specified amount of angular momentum or spin to a solution of the vacuum Einstein equations, producing a new solution with specified angular momentum or spin but with only slightly perturbed energy-momentum vector for asymptotic flat spacetimes.  

In this article, we have shown the existence of the conserved quantities in terms of well-posed partial differential equations in a generic curved spacetime even with no Killing vectors. Conserved charges discussed in this article, are not associated with diffeomophism invariance of spacetime or in other words Killing vector fields. Existence of these conserved charges are independent of asymptotic structure of spacetime manifold. However, like Killing equations, there also exist spacetime manifolds for which solutions of those partial differential equations do not exist and as a result, these conserved charges don't exist for dynamical systems in those spacetime manifolds. Furthermore, we also discuss the closed algebra between these conserved charges under the Poisson bracket both in free-particle case and also with non-minimal spin-curvature coupling. It is shown that these conserved charges for a given dynamical system in curved spacetime, form an infinite-dimensional Lie-algebra similar to the Virasoro algebra. 

\section{Acknowledgement}
SM wants to thank CSIR to support this work through doctoral fellowship.

\bibliographystyle{unsrt}
\nocite{*}
\bibliography{draft}

\end{document}